\documentclass[aps,prd,nofootinbib,superscriptaddress,showpacs,floats,floatfix,notitlepage,10pt]{revtex4-1}

\usepackage{graphicx}
\usepackage{amssymb}
\usepackage{amsmath}
\usepackage{amsfonts}

\usepackage{bm}
\usepackage{mathrsfs}
\usepackage[colorlinks=true]{hyperref}
\usepackage[all]{hypcap} 
\usepackage[utf8]{inputenc} 
\usepackage{mathptmx}      
\usepackage{color}
\usepackage{multirow}
\usepackage{palatino}
\usepackage{journals}
\usepackage{soul}
\usepackage{calrsfs}
\DeclareMathAlphabet{\pazocal}{OMS}{zplm}{m}{n}
\usepackage{amsthm}
\newtheorem{thm}{Theorem}[section]
\newtheorem{prop}{Proposition}[section]

\newcommand{\beq}{\begin{equation}}
\newcommand{\eeq}{\end{equation}}

\newcommand{\bse}{\begin{subequations}}
\newcommand{\ese}{\end{subequations}}

\newcommand{\ain}[1]{$\boldsymbol{#1}$}
\newcommand{\ainF}[1]{\boldsymbol{#1}}

\newcommand{\ie}{i.e.,~}
\newcommand{\eg}{e.g.,~}

\begin{document}
\title{
Preserving Kerr symmetries in deformed spacetimes}

\author{Georgios O. Papadopoulos}
\email{gop@disroot.org}
\affiliation{Department of Physics, Aristotle University of Thessaloniki, 54124, Greece}
\author{Kostas D. Kokkotas}
\email{kostas.kokkotas@uni-tuebingen.de}
\affiliation{Theoretical Astrophysics, IAAT, Eberhard-Karls University of T\"{u}bingen, 72076 T\"{u}bingen, Germany}
\date{\today}
\begin{abstract}
We present a novel approach in constructing deviations of the Kerr spacetime whereas the symmetries can be preserved. 
The method was applied trivially in all known classical black-hole spacetimes tested, while provides the possibility of testing and inventing deformations of Kerr in a quick and innovative way. 
The methodology is based on earlier work by  Benenti and Francaviglia \cite{Benenti1979} on separability of geodesic equations while we exploited the 
elegant properties of Yano tensor in order to provide an additional way of getting the Carter constant. 
\end{abstract}

\maketitle
%
\section{Introduction}

Black holes play an important role in our understanding of the Universe and fundamental interactions \cite{frolov2011introduction}. From the point of view of the classical field theory, they are apparently the simplest objects ever known, because they can be fully described by two parameters, mass and angular momentum (and, in cases of non-vacuum theories, also by the corresponding charges). Black holes are essential for the formulation of the renown gauge/gravity duality \cite{2014Sci...344..806M}, allowing one to investigate the strongly coupled fields theories, as well as to get an insight into quantum gravity.

The great leap made during recent years in astronomy has brought the black hole topic from the theoretical area to an experimental one. The next generation of X-ray missions will be able to detect with unprecedented accuracy the spectrum of the radiation generated from the inner part of the accretion disk~\cite{2016SPIE.9905E..1QZ}. Observation of the shadow cast by supermassive black holes might be possible though the international effort consisting in combining data from various radio-telescopes (the Event Horizon Telescope) \cite{2014ApJ...788..120L,2015Sci...350.1242J}. More recently the sensational observations of gravitational waves, by the LIGO and VIRGO collaborations, from the merger of two black holes \cite{2016PhRvL.116x1103A,2016PhRvL.116v1101A,2016PhRvL.116f1102A} opened a new era in astronomy, astrophysics, and gravitational theory.

In the weak field regime Einstein's gravity was confirmed up to a few orders of post-Newtonian expansion. The detection of gravitational waves \cite{2016PhRvL.116x1103A,2016PhRvL.116v1101A,2016PhRvL.116f1102A} is the first test of strong gravity regime that cannot be effectively modelled only within the post-Newtonian approach. Although Einstein's theory is consistent with the experiment, the current, rather large, indeterminacy in the measurement of the black hole mass and angular momentum leaves also the window for alternative theories open \cite{2016PhRvL.116v1101A, 2016PhRvL.116q1101C, 2016PhLB..756..350K}. According to the planned increasing of the accuracy in the advanced LIGO \cite{2016PhRvL.116v1101A}, the parametric indeterminacy cannot be completely constrained and the window for alternative theories should remain open.

 An opportunity to measure possible deviations from Einstein's gravity is to test the geometry of black hole event horizon with electromagnetic radiation or by analysing the quasi-normal mode ringing of the newly created black-holes \cite{2016PhRvL.116v1101A,2017ApJ...842...76B,2017RvMP...89b5001B,2018PhRvL.120e1101C}. This can be achieved by constraining possible deviations from the exact solution of the vacuum Einstein equations, representing axially symmetric, asymptotically flat black hole - the Kerr solution \cite{1963PhRvL..11..237K}. The Einstein theory, being one of the simplest among existing theories of gravity, leads to quite complicated equations for arbitrary axially symmetric space-time and, usually, an Ansatz is necessary for finding a new exact solution of this kind. Therefore, it is not surprising that no much progress has been made in finding other axially symmetric black hole solutions during more than fifty years after Kerr's original work \cite{1963PhRvL..11..237K}.

The list of exact solutions for four-dimensional axially symmetric black holes is limited by a few Kerr-like space-times with extra parameters, which include electric and/or magnetic charges \cite{1965JMP.....6..918N, 1992PhRvL..69.1006S, 1986AnPhy.172..304M, 2006CQGra..23.5323C}. These known generalisations of the Kerr solution mostly are not interesting for astrophysics, because the value of the electric charge is negligibly small, while the magnetic charge has not been observed at all. The Kerr metric (in Boyer -- Lindquist coordinates), is Petrov-type D and admits four constants of motion \ie the particle rest mass $\mu$, its energy $E$, the angular momentum about the $z$-axis $L_z$, and the Carter constant.

In this way, we have a situation in which there are a number of alternative theories of gravity, emanating from the need to answer fundamental questions related to the hierarchy problem, the dark energy and dark matter problem, the nature of curvature singularities, the construction of quantum gravity, etc. These alternative theories produce the same post-Newtonian behaviour, but, when considering the strong gravity regime near the black hole event horizon,  no exact solutions for astrophysically relevant rotating black holes are known except the one in the Einstein theory - the Kerr solution.
Thus, the starting point for testing alternative theories of gravity should be constructing viable models for a rotating black hole in these theories and using a  way of measuring the potential deviations \cite{2017PhRvD..96f4054G} in the spirit of the PPN approximation.
By now, due to the above complexity in finding new axially symmetric exact solution, there are only approximate black hole metrics obtained either perturbativelly in the regime of slow rotation and weak coupling or numerically in a few theories of gravity, such as dynamical Chern-Simons or Einstein-dilaton-Gauss-Bonnet ones \cite{2011PhRvD..83j4002Y,2014PhRvD..90d4066A,2009PhRvD..79h4043Y, 2016PhRvD..93d4047K}. The perturbative metrics cannot be applied to highly or even moderately rotating black holes and considerable values of coupling constants.

A number of spacetimes deviating from Kerr has already been proposed for testing the ``no-hair theorem'' via observations in both electromagnetic and gravitational wave spectra \cite{2006CQGra..23.4167G,2010ApJ...716..187J,2010ApJ...718..446J,2013ApJ...773...57J,2014ApJ...784....7B,2014PhRvD..89f4007C,2016PhRvD..93f4015K}. The majority of them is very much similar to the Kerr at the weak field or/and slow rotation limit but there may present significant deviations in the strong field regime. These somehow ad-hoc created metrics  may endowed by two commuting Killing fields but they do not admit the ``fourth'' constant which characterises the Kerr spacetime \ie the Carter constant.  Carter's constant \cite{1968CMaPh..10..280C,2009GReGr..41.2873C} is extremely important for the determination of the observable quantities since they simplify significantly (separate) the equations of motion. Actually the Hamilton-Jacobi equations are  separable in all four coordinates making the geodesic equations integrable. Johannsen \cite{2013PhRvD..88d4002J} designed a Kerr-like black hole metric which is regular outside the horizon, has {three} constants of motion and depends on four extra functions characterising the deformation.
 In early 2018, Konoplya et al.\cite{PhysRevD.97.084044} presented a class of axisymmetric and asymptotically flat black-hole spacetimes for which the Klein-Gordon and Hamilton-Jacobi equations allow for the separation of variables. 
More recently, Glampedakis and Pappas \cite{2018arXiv180604091P,2018arXiv180609333G} discussed the spherical photon orbits as a diagnostic of non-Kerr spacetimes.

Here we provide a criterion for testing if a given spacetime admits  a ``fourth'' constant. This criterion originally was tested for trivial spacetime cases  and subsequently it was used to examine the existence of the Carter integral for the two most popular deformations of Kerr spacetime.  These are the ``quasi-Kerr'' case \cite{2006CQGra..23.4167G} and the metric designed by Johannsen \cite{2013PhRvD..88d4002J}. 

The paper is organised as follows: in Section II we give an invariant proposal on how to 
construct alternative geometries for axisymmetric spacetimes admitting which are black-hole solutions (which one could call ``Kerr
deformations'') under the fundamental assumption of the preservation of the Kerr features like the
Carter constant and the asymptotic flatness.
In subsection III we apply our scheme to some interesting cases. Initially,  we test the simple cases (Kerr, Kerr-Newman, Kerr-Sen,\dots) known to admit a ``fourth'' constant. Then we examine the ``quasi-Kerr'' case \cite{2006CQGra..23.4167G} and the metric suggested by Johannsen \cite{2013PhRvD..88d4002J}.

The paper concludes with a discussion section and two Appendices; in the first we present the theorem on which the technique presented here is based, while in the second a short study on the existence and integrability of Yano tensors, and  their relation to the Killing tensors is presented, thus putting the mathematical foundations for the statement of the invariant proposal.

\section{Alternative black hole geometries: an invariant proposal}
\label{sec:proposal}

Right from the outset it will be assumed that any alternative geometry meant to describe a physical black hole, must share at least the properties of the Kerr metric; in other words the two commuting Killing vector fields (one time-like and one space-like) and the notion of the Carter constant must be preserved. From the physical point of view, this preservation is tantamount to the separation of those Hamilton-Jacobi equations which correspond to the geodesic equations for a test particle
\beq\label{HJ}
g^{ij}\partial^{}_{i}S\partial^{}_{j}S=\text{const.} \in \mathbb{R}
\eeq
Indeed, the separability is closely connected with the existence of a fourth (quadratic in momenta) integral of motion \cite{Eisenhart1997,Benenti1979}  which is identified as the Carter constant; the first integral is the invariant proper mass, the second is the energy, {and} the third is a component of the angular momentum of the test particle.

Using {a theorem} provided by Benenti and Francaviglia \cite{Benenti1979,zbMATH03710954,MR583725} (see detailed presentation in Appendix \ref{sec:AppendixA})  and  after a thorough analysis, one arrives at the following result: there exists a local coordinate system $(r,x,\phi,t)$ of the Boyer-Lindquist type, compatible with the separability of the Hamilton-Jacobi \eqref{HJ} (with $n=2$ in the theorem to be found in the Appendix A), such that the contravariant form of the underlying metric tensor assumes the form
\beq
g^{ab}_{}=\frac{1}{A_{1}(r)+B_{1}(x)}
\begin{pmatrix}
A_{2}(r) & 0 & 0 & 0\\
0 & B_{2}(x) & 0 & 0 \\
0 & 0 &A_{3}(r)+B_{3}(x) & A_{4}(r)+B_{4}(x)\\
0 & 0 &A_{4}(r)+B_{4}(x) & A_{5}(r)+B_{5}(x)
\end{pmatrix}
\label{eq:con_metric}
\eeq
while the corresponding (non trivial) contravariant Killing tensor (of rank two) reads:
\beq
K^{ab}_{}=\frac{1}{A_{1}(r)+B_{1}(x)}
\begin{pmatrix}
A_{2}(r)B_{1}(x) & 0 & 0 & 0\\
0 & -B_{2}(x)A_{1}(r) & 0 & 0 \\
0 & 0 &B_{1}(x)A_{3}(r)-A_{1}(r)B_{3}(x) & B_{1}(x)A_{4}(r)-A_{1}(r)B_{4}(x)\\
0 & 0 &B_{1}(x)A_{4}(r)-A_{1}(r)B_{4}(x) & B_{1}(x)A_{5}(r)-A_{1}(r)B_{5}(x)
\end{pmatrix}
\eeq
Obviously, the metric tensor field admits the symmetries of the Kerr black hole, i.e., the two Killing vector fields related to the axial symmetry.

This theorem is an extremely useful tool both in testing proposed solutions but also in trying to construct solutions that deviate from Kerr but still share its symmetries. Moreover, using this theorem and the invariant 
proposal one can give to the notion of \emph{asymptotic symmetry} a concise meaning: \emph{a Kerr deformation is asympoticaly endowed with a Carter constant if and only if at large distances $(r\rightarrow\infty )$ assume the prescribed form.}

Now we are in position to state the invariant proposal:
\begin{prop}
Any alternative geometry meant to describe the physics of a black hole, deviating smoothly from the Kerr black hole, must be
\begin{itemize}
	\item[(i)] susceptible either to an algebraically general Yano tensor of rank two, in which the two distinct eigenvalues supposed to be non constant, or to a non trivial Killing  tensor of rank two along with two non null, and commuting, Killing vector fields;
	\item[(ii)] asymptotically flat .
\end{itemize}
\end{prop}
The next section is devoted to a few examples to be found in the relevant literature, all indicating that they comply with our proposal.

\section{Applications}

\subsection{Testing the classical cases}
\label{sec:2A}

Here we examine how the central statement, defined earlier, is applied to a list of classical examples. In the literature, there exist various black hole geometries ; some from the realm of General Relativity \cite{1973grav.book.....M} (like, Schwarzschild, Reissner-Nordstr\"om, Kerr and Kerr-Newman) and others from alternative theories like Strings (e.g., the Modified Kerr \cite{2016PhLB..756..350K}, and the Sen \cite{1992PhRvL..69.1006S}). 

All these metrics can be unified in a single 6-parametric family. Indeed in a local coordinate system, which is part of the atlas of the manifold, say that of Boyer-Lindquist
$x^{a}=(r,x,\phi,t)$, ($\cos{\theta}=x$) one can choose the following coframe:
\begin{equation}
\ainF{\omega}^{1}=\sqrt{\frac{\rho}{\Delta}}dr \, ,\quad
\ainF{\omega}^{2}=\sqrt{\frac{\rho}{1-x^{2}}}dx\, ,\quad
\ainF{\omega}^{3}=\sqrt{\frac{1-x^{2}}{\rho}}(a dt-f d\phi)\, ,\quad
\ainF{\omega}^{4}=\sqrt{\frac{\Delta}{\rho}}(dt-a(1-x^{2})d\phi)
\end{equation}

and the rigid metric
\beq
\eta_{ab}=
\begin{pmatrix}
1 & 0 & 0 & 0\\
0 & 1 & 0 & 0\\
0 & 0 & 1 & 0\\
0 & 0 & 0 & -1
\end{pmatrix}
\eeq
where
\begin{eqnarray}
\label{eq:Delta}
\Delta&=&r^{2}+a^{2}-2r(M-b)+\varepsilon Q^{2}-H/r,\\
\rho&=&r^{2}+a^{2}x^{2}+2br\\
f&=&r^{2}+a^{2}+2b r
\label{eq:f(r)}
\end{eqnarray}
so that the line element reads $\ainF{g}=\eta^{}_{ab}\ainF{\omega}^{a}_{}\cdot\ainF{\omega}^{b}_{}$.
The various constants entering in the line elements are, mass ($M$), angular momentum per unit mass ($a$), charge ($Q$) and some magnetic multiple moment (say $b$ in the simplest case) per unit mass. 

Then this line element describes
\begin{enumerate}
\item the Kerr-Newman black hole when only $M\neq0$, $a\neq0$, $\varepsilon=1$ and $Q\neq0$
\item the modified Kerr black hole when only $M\neq0$, $a\neq0$ and $H\neq0$
\item the (Kerr-)Sen black hole when only $M\neq0$, $a\neq0$, $b\neq0$ 
\end{enumerate}

The entire 6-parametric family admits metrics which can be brought in the form given by equation (\ref{eq:con_metric}) and complies with the conditions of the theorem \ref{TheoremI}. Indeed, one can verify that the allocations
\bse
\begin{align}
&A_{1}(r)= r^{2}+ 2 b r  \\
&A_{2}(r)=\Delta \\
&A_{3}(r)=-\frac{a^2}{\Delta} \\
&A_{4}(r)=-\frac{f}{\Delta}a\\
&A_{5}(r)=-\frac{f^2 }{\Delta} \\
&B_{1}(x)=a^{2} x^{2}\\
&B_{2}(x)=1 - x^{2}\\
&B_{3}(x)=\frac{1}{1-x^{2}}\\
&B_{4}(x)=a\\
&B_{5}(x)=a^{2}(1-x^{2})
\end{align}
\ese
identify the family with the prescribed metric.
It is not hard to see (with e.g., the use of a Computer Algebra System) that the first two metrics, which are of Petrov type $D$, are susceptible to (the same) algebraically general Yano tensor (in coframe components) 
\beq
Y^{(1,2)}_{ab}=\begin{pmatrix}
0 & 0 & a x & 0\\
0 & 0 & 0 & r\\
-a x & 0 & 0 & 0\\
0 & -r & 0 &0
\end{pmatrix} \quad \Rightarrow \quad
K^{(1,2)}_{ab}=\begin{pmatrix}
-a^{2}x^{2} & 0 & 0 & 0\\
0 & -r^{2} & 0 & 0\\
0 & 0 &a^{2}x^{2}  &0\\
0 & 0 & 0 & -r^{2}
\end{pmatrix}, 
\eeq
or
\beq
K^{(1,2)}_{ab}=Y^{(1,2)}_{am}\eta^{mn}_{}Y^{(1,2)}_{nb},
\eeq
which, in turn, gives rise to the two manifest Killing vector fields (along with the induced Killing tensor of order two). 

The third metric (Sen), which is of Petrov type $I$, does not admit a Yano but rather a Killing tensor \cite{2008PhRvD..78d4007H} along with one (essentially, two) manifest Killing vector fields (coframe components)
\beq
K^{(3)}_{ab}=\begin{pmatrix}
-a^{2}x^{2} & 0 & 0 & 0\\
0 & -2 b r-r^{2} & 0 & 0\\
0 & 0 &a^{2}x^{2}  &0\\
0 & 0 & 0 & -2 b r-r^{2}
\end{pmatrix} \, .
\eeq
Obviously, at large distances ($r\rightarrow\infty$ or, equivalently when $b\rightarrow 0$) then $K^{(3)}_{ab}\rightarrow K^{(1,2)}_{ab}$, and asymptotically the Kerr-Sen metric admits
a Yano 2-form.

At this stage, and under the point of view of the theorem, one could ask: \emph{what is the usefulness of the Yano tensor?} The answer to this reasonable question is threefold. On one side one can construct deformations which preserve the form of the Yano tensor (and thus the form of the induced Killing tensor as well). For instance, the Modified Kerr solution \cite{2016PhLB..756..350K} is gained by a simple transformation
\beq
M\rightarrow M+\frac{H}{2r}
\eeq
a transformation which leaves the Yano tensor invariant. On the other hand a simpler (and thus, coarse) classification of those deformations which are based on the existence of a Yano tensor is permitted. Finally, one could prescribe a Yano tensor for a hypothetical model and, by using the equations found in the Appendix B, one could estimate the features of the compatible (to the prescribed Yano) Ricci tensor, and thus, via some related field equations, the corresponding energy momentum tensor.

\subsection{The quasi-Kerr spacetime}
\label{sec:2B}

The effort to test the no-hair theorem led to proposal which either involved spacetime perturbations   \cite{PhysRevD.69.124022} or a multipolar expansion \cite{PhysRevD.52.5707,1992CQGra...9.2477M}. In this spirit
Glampedakis \& Babak \cite{2006CQGra..23.4167G}
proposed the so called ``quasi-Kerr'' spacetime by using the Hartle-Thorne \cite{1967ApJ...150.1005H,1968ApJ...153..807H} formalism to add terms describing a slowly rotating body. Their formalism  provides an expansion up
to the quadrupole order and has a quadrupole moment $Q$ that is independent of both mass and spin. 
This metric is stationary, axisymmetric and asymptotically flat whilst it reduces to the Kerr metric in the absence of deformation. 
The quasi-Kerr metric used extensively over in the last decade in most of the studies related to astrophysical test of the non-hair theorem and in general testing of the Kerr solution.

In relation to our study, the quasi-Kerr metric  admits separability of the Hamilton-Jacobi equation only for equatorial orbits. For arbitrary orbits, the Carter constant cannot be found and the only remaining way is the numerical intergration \cite{2005MNRAS.358..923B}. It is not difficult, following the conditions set in the previous section, to verify the last statement \ie the absence of a Carter constant.

\subsection{A realistic example: the Johannsen metric}
\label{sec:2C}

The Johannsen metric \cite{2013PhRvD..88d4002J}, constructed to admit a Carter constant, obviously complies with the conditions of the theorem. Indeed, a tedious yet straightforward procedure shows that the needed allocations are
\bse
\begin{align}
&A_{1}(r)=r^2+ \mathcal{F}(r) \\
&A_{2}(r)=\Delta(r) \mathcal{A}_{5}(r)\\
&A_{3}(r)=-\frac{a^{2}}{\Delta(r)} \mathcal{A}_{2}(r)^{2}\\
&A_{4}(r)=-\frac{f}{\Delta}a\mathcal{A}_{1}(r)\mathcal{A}_{2}(r)\\
&A_{5}(r)=-\frac{f^2}{\Delta(r)} \mathcal{A}_{1}(r)^2\\
&B_{1}(x)=a^2 x^2\\
&B_{2}(x)=1-x^2\\
&B_{3}(x)=\frac{1}{1-x^2}\\
&B_{4}(x)=a\\
&B_{5}(x)=a^2 (1-x^2)
\end{align}
\ese

where the deviation functions $\mathcal{A}_{1}(r)$, $\mathcal{A}_{2}(r)$, $\mathcal{A}_{5}(r)$ and $\mathcal{F}(r)$ were chosen to be described in terms of power series of $M/r$ by Johannsen \cite{2013PhRvD..88d4002J}. The functions $\Delta$ and $f$ are described via the equations (\ref{eq:Delta}) and (\ref{eq:f(r)}).

So, the main examples in the literature obey all the conditions of the suggested invariant proposal.

The similarity in the structure of the classical metrics in subsection \ref{sec:2B} with the structure invented by Johannson \cite{2013PhRvD..88d4002J} indicates that there is a simple way in generating deviations to the known metrics whilst preserving their main symmetries. By following the approach suggested here one can avoid the ingenious but still quite laborious effort of Johannson in constructing this type of spacetimes.

It should be underlined the fact that the metric tensor field \eqref{eq:con_metric} is the most general one can have. It includes
all the Petrov type D solutions (like Kerr, modified Kerr, etc.) but  also the Petrov type I as well (\eg Sen metric).
On the other hand the Johannsen metric is just a sub case of the proposed metric. Indeed there no proper allocations for the parameters of the Johannsen metric to include the Sen metric. Never the less, it seems that the Johannsen metric is the most general one when it comes to the type D solutions. Of course such an estimation needs verification --something which is beyond the goals of the present paper, especially from the point of view of the generality of the proposed solution.

\section{Discussion}

In the literature, one can find many alternative models to general relativistic black holes (most prominently, of the Kerr BH). Nevertheless, it is the physics that poses a minimum of requirements governing these models. For instance, the preservation of axial symmetry and the separation of the geodesic equations seem to comprise that minimum. In the present paper, based on that minimum of physical requirements, we present a concise treatment of this problem by
\begin{itemize}
	\item giving a precise meaning to the notion of ``Kerr deformation'',
	\item proposing an invariant criterion which not only allows to check the viability of one's model but also to 
	      construct alternative models endowed ab initio with the minimum of physical properties,
    \item giving a precise meaning to the notion of ``asymptotically symmetric model'',
	\item in some cases (i.e., when a Yano tensor is permitted) we give a neat way of a (coarse) classification of the model which, at the same time, offers a naive way for proposing models.
\end{itemize}

This some how technical work will be followed with applications in astrophysics \ie in constructing wave equations under these very general conditions, applications to the studies of black-hole shadows and iron lines.
\vskip 1cm

\acknowledgements{This work was supported by the DAAD program  ``Hochschulpartnerschaften mit Griechenland 2016'' (Projekt 57340132). The authors are grateful to  C. Bambi, R. Konoplya, S. Nampalliwar, S. Yazadjiev  and A. Zhidenko for useful comments that improved the article.}

\appendix

\section{The Theorem upon which the proposal is based}
\label{sec:AppendixA}
Based on a simple train of thoughts one arrives at the proposal of Section  \ref{sec:proposal}. Its purpose is to describe the physics of a black hole, deviating smoothly from the Kerr black hole. This is not a trivial task, but luckily Benenti and Francaviglia \cite{Benenti1979,MR583725,zbMATH03710954} proved the following, extremely useful, theorem:
\begin{thm} \label{TheoremI}
A manifold $(\mathcal{M},\ainF{g})$ admits a local separability structure, for the Hamilton-Jacobi system \eqref{HJ}, if and only if the following conditions hold:
\begin{itemize}
	\item[(i)] there exist (locally) $n$ independent, commuting Killing vectors $\ainF{X}_{(a)}, 1 \leq a\leq n$
	\item[(ii)] there exist (locally) $4-n$ independent, commuting (\`a la Schouten–Nijenhuis) Killing tensors $\ainF{K}_{(a)}$ which are annihilated by the Killing vectors
	\item[(iii)] the Killing tensors have in common $4-n$ commuting eigenvectors such that they are perpendicular to the Killing vectors and also annihilated by them.
\end{itemize}	
\end{thm}

\section{On Yano and Killing tensors of rank two}
\label{sec:AppendixB}

Here we prove that the   existence of an algebraically general Yano tensor of rank two, with non constant
eigenvalues leads to the existence of two commuting Killing vector fields and the Carter's constant.

\medskip

The very existence of a Carter constant presupposes the existence of a Killing tensor (the simplest being a tensor of rank two). 
Now, there are two cases for that Killing tensor; either it is induced by a Yano tensor (as its `square') or it is not.
In any case, the susceptibility to a Killing tensor is closely related to very interesting properties like the separation of variables for various relative 
(to the geometry) partial differential equations, \eg the Hamilton-Jacobi systems etc. \cite{Eisenhart1997}.
In the second case, the Carter constant is given from the beginning, while in the first case things are less trivial. Indeed, 
let an algebraically general Yano tensor of rank two. Its matrix rank will be four, while the two defining eigenvalues 
are supposed to be non constant. The assumed Yano tensor corresponds to a 2-form satisfying the defining relation
\beq\label{Yano_Definition}
\nabla_{(c}Y_{b)a}=0, ~~~\text{and} ~~~Y_{(ab)}=0 \, .
\eeq
Its is obvious that the previous definition renders the first covariant derivative of the Yano tensor, 
which corresponds to a 3-form, completely antisymmetric; thus it can be seen as the Hodge dual of a 1-form, 
say, $\ainF{X}$:
\beq\label{First_Vector}
\nabla^{}_{c}Y^{}_{ab}=\varepsilon^{}_{abcm}X^{m}_{}
\eeq
or in exterior calculus notation $d\ainF{Y}=\star\ainF{X} $
which, since $d^{2}(\ldots)=0$, the previous relation implies that $d\star\ainF{X}=0$.
In other words, the covariant divergence of the corresponding (co)vector vanishes ($\nabla\cdot\ainF{X}=0)$.
In view of \eqref{First_Vector} one can get an alternative (yet equivalent) definition for the Yano 2-form
\beq\label{Star_Yano_Definition}
\nabla^{}_{c}(\star Y)^{ab}_{}=2 X^{[a}\delta^{b]}_{c} \, .
\eeq

Using the Ricci identities and the definition of the Yano tensor \eqref{Yano_Definition} one can get the following 
integrability conditions:
\beq\label{Differential}
2\nabla^{}_{d}\nabla^{}_{c}Y^{}_{ab}=2Y^{}_{ma}R^{m}_{\phantom{1}dbc}+Y^{}_{mb}R^{m}_{\phantom{1}acd}
+Y^{}_{mc}R^{m}_{\phantom{1}adb}+Y^{}_{md}R^{m}_{\phantom{1}acb}
\eeq
which, in turn, result in 
\begin{eqnarray}\label{Algebraic1}
&Y^{}_{ma}(R^{m}_{\phantom{1}dbc}+R^{m}_{\phantom{1}cbd})+Y^{}_{mb}(R^{m}_{\phantom{1}dac}+R^{m}_{\phantom{1}cad})
+Y^{}_{mc}(R^{m}_{\phantom{1}adb}+R^{m}_{\phantom{1}bda})+Y^{}_{md}(R^{m}_{\phantom{1}acb}+R^{m}_{\phantom{1}bca})=0
\end{eqnarray}
while a further contraction gives
\beq\label{Algebraic2}
Y^{}_{ma}R^{m}_{\phantom{1}b}+Y^{}_{mb}R^{m}_{\phantom{1}a}=0 \, .
\eeq
Moreover, in a similar fashion using the Ricci identities, the definition of the Hodge dual Yano tensor 
\eqref{Star_Yano_Definition} and the fact that $\nabla\cdot\ainF{X}=0$, one can get the following integrability 
conditions:
\beq\label{First_Killing}
2\nabla^{}_{(b}X_{a)}=-(\star Y)^{}_{(a|m|}R^{m}_{\phantom{1}b)} \, .
\eeq

An immediate consequence of the property of algebraic generality and equation \eqref{Algebraic2} is that the Ricci tensor 
assumes such a form that \ain{X} is a (real), as per of equation \eqref{First_Killing}, Killing vector field.

This is the proper point in the analysis, to ask about the behaviour of the Yano tensor under the Lie dragging with 
respect the Killing vector field \ain{X}. First of all, a key observation: a tedious yet straightforward calculation 
can show that the Lie derivative with respect a Killing field and the covariant derivative, as operations, commute 
-- so in particular it is
\beq
[\pounds_{\ainF{X}},\nabla_{a}](\text{any tensor})
\overset{\ainF{X}}{\underset{\text{KVF}}{=}}0 \, .
\eeq
Thus
\beq
[\pounds_{\ainF{X}},\nabla_{a}]Y_{bc}=0
\eeq
which, in view of \eqref{First_Vector} implies that
\beq
\nabla_{a}\pounds_{\ainF{X}}Y_{bc}=0
\eeq
i.e., the Lie derivative of the Yano tensor with respect the vector \ain{X} is covariantly constant. Let
\beq
\pounds_{\ainF{X}}Y_{bc}=C_{bc}
\eeq
where $C_{bc}$ is an antisymmetric, covariantly constant tensor.
Now taking the Lie derivative, with respect \ain{X}, of the \eqref{Differential} ---in view of the previous points--- 
implies that the components of $C_{bc}$ all vanish, thus
\beq\label{Lie_Dragging_Yano}
\pounds_{\ainF{X}}Y_{bc}=0 \, .
\eeq

It is natural to examine whether there are further, induced and based on the Yano tensor, symmetries. The simplest possible candidate is
the vector field
\beq\label{Second_Vector}
H^{a}_{}=Y^{a}_{\phantom{1}b}Y^{b}_{\phantom{1}c}X^{c}_{}
\eeq
for, it is simpler to consider quantities which depend only on the Yano tensor rather than on its covariant derivatives.
Never the less, one can read off from the previous definition a Killing tensor of rank two, i.e.,
\beq\label{Killing_Tensor}
K^{}_{ab}=Y^{}_{am}\eta^{mn}_{}Y^{}_{nb}, ~~~\text{and} ~~~\nabla_{(a}K_{bc)}=0
\eeq
The next natural question is about  the Lie dragging, with respect the Killing vector field \ain{X}, of the Killing tensor.
On one hand, since \eqref{Lie_Dragging_Yano} and \eqref{Killing_Tensor} hold one concludes that
\beq\label{Lie_Vanishing_Killing_T}
\pounds_{\ainF{X}}K_{ab}=0
\eeq
on the other hand, using the very definition of the Lie derivative for a covariant tensor of rank two and the definition \eqref{Second_Vector}, one arrives at
\beq
\pounds_{\ainF{X}}K_{ab}=2\nabla_{(a}H_{b)}
\eeq
so
\beq
\pounds_{\ainF{H}}\eta_{ab}=0
\eeq
i.e., there is yet another Killing vector field.
Finally, due to \eqref{Lie_Vanishing_Killing_T}, one can easily prove that the two Killing vector fields commute.

\emph{Thus, the initial assumption on the existence of an algebraically general Yano tensor of rank two, with non constant
eigenvalues leads to the existence of two commuting Killing vector fields and the Carter's constant -- via the induced, by the
Yano tensor, of a Killing tensor of rank two.}

The very existence of a Killing tensor of order two leads to the Carter constant; an integral of the geodesics. Indeed, let $\ainF{U}$ be a geodesic field with affine parameter $\tau$
\beq
U^{b}_{}\nabla^{}_{b}U^{a}_{}=0
\eeq
then the quantity $C\equiv K^{}_{ab}U^{a}_{}U^{b}_{}$
remains constant along the geodesic curve, since
\beq
\pounds_{\ainF{U}}C=\pounds_{\ainF{U}}(K^{}_{ab}U^{a}_{}U^{b}_{})=\nabla^{}_{(c}K^{}_{ab)}U^{a}_{}U^{b}_{}U^{c}_{}
+K^{}_{cb}(\nabla^{}_{a}U^{c}_{})U^{a}_{}U^{b}_{}+K^{}_{ca}(\nabla^{}_{b}U^{c}_{})U^{a}_{}U^{b}_{}
{=}0
\eeq

where we used the equations  $\nabla^{}_{(c}K^{}_{ab)}=0$ and ${U^{b}_{}\nabla^{}_{b}U^{a}_{}=0}$.

To put it differently, the geodesic equations (in coordinate space)
\beq
\ddot{x}^{a}+\Gamma^{a}_{\phantom{1}bc}\dot{x}^{b}_{}\dot{x}^{c}_{}=0, ~~~\text{where} ~~~\dot{x}^{a}_{}\equiv\frac{d x^{a}_{}}{d\tau}
\eeq
admit an integral of a second order
\beq
C\equiv K^{}_{ab}\dot{x}^{a}_{}\dot{x}^{b}_{}, \quad \mbox{with} \quad {\dot C}=0 
\eeq
if and only if 
\beq
\nabla^{}_{(c}K^{}_{ab)}=0
\eeq
the proof being straightforward. Of course, the same result holds for the (equivalent to the geodesics equations) corresponding Hamilton-Jacobi equation.

%
\bibliography{references}
%

\end{document}